\documentclass[a4paper]{jpconf}
\pdfoutput=1
\usepackage{graphicx}
\usepackage{amsmath}
\usepackage{pstricks}
\usepackage{float} 
\usepackage[charter,expert]{mathdesign}

\usepackage{cite}
\usepackage[%
  unicode=true,%
  colorlinks=true,%
  linkcolor=blue,anchorcolor=blue,%
  breaklinks=true,%
  citecolor=blue,filecolor=cyan,%
  menucolor=blue,%
  urlcolor=blue]{hyperref}

\begin{document}
\title{Energy spectra of a spin-$\tfrac12$ XY spin molecule
interacting with a single mode field cavity: Numerical study}

\author{H~Tonchev,\ \ A~A~Donkov,\ \ and\ \ H~Chamati}

\address{Institute of Solid State Physics,
Bulgarian Academy of Sciences,

72 Tzarigradsko Chauss\'ee, 1784 Sofia, Bulgaria}

\ead{htonchev@issp.bas.bg, aadonkov@issp.bas.bg, hassan.chamati@issp.bas.bg}

\begin{abstract}

In a previous paper 
[\href{http://dx.doi.org/10.1088/1742-6596/682/1/012032}{J. Phys.: Conf. Ser. 682 (2016) 012032}] we studied analytically the energy spectra of a finite-size spin $\tfrac12$ XY chain (molecule) coupled at an arbitrary spin site to a single mode of an electromagnetic field via the Jaynes-Cummings model. We considered spin rings and open spin molecules with up to 4 spins and an interaction restricted to nearest-neighbours. Here we extend our investigation, addressing numerically the energy spectra of molecules of up to 10 spins with nearest-neighbour or long-range interaction. Furthermore we analyze the behaviour of an invariant operator, constructed by combining the magnetization of the spin-chain and the total number of photons in the system. We found a strong dependence on the number (even or odd) of sites in the molecules. This study is aimed at finding the appropriate combination of the physical parameters that could make the system suitable for use in quantum computations.

\end{abstract}

\section{Introduction}
The Jaynes-Cummings~\cite{EJFC1963,CGPK2005} model for the interaction of an atom with a quantized electromagnetic field has been used to investigate atom- and ion-field interactions in an ion trap and quantum dots, to study one photon lasers, and for single-photon photodetectors, see for example Refs.~\cite{WVRM1995,JYFN2003,GRHW1987,ADSM2006}, respectively. Here this model is used in a numerical calculation to obtain the energy spectrum of a magnetic spin chain, or more in tunes with this conference, a magnetic nano-material in the form of a spin molecule, interacting with a laser field. 

We consider the easiest, from the standpoint of numerical computation, case of $\tfrac12$ spins of the individual sites in a molecule. The spin-spin interaction has been modeled by the XY model~\cite{JBMC2010}, and cases of nearest neighbour and long range interactions are considered. In a previous study~\cite{Tonchev:2016JPCS} we had shown some analytical formulae for the energy spectrum for a 4-site spin molecule, with the numerics we were able to go up to 10-site molecules, but it becomes hard to visualize the full spectra. Due to the specifics of the coupling between the Jaynes-Cummings model and the product states, that diagonalize the spin-spin Hamiltonian, the energy spectrum shows qualitative differences whether we have an odd or an even number of spins in the spin molecule.

In Section~\ref{sec:Hamiltonian} we remind the general form of the interaction Hamiltonian used and show some of the spectra for different chain geometries. Then we analyze the obtained spectra in Section~\ref{sec:analyze}. We look into the analogy between the invariant that exists in this problem, which helps in classifying the energy levels, with the total magnetization in Section~\ref{subsec:Invariant}, then consider the influence of the longer range interaction in Section~\ref{subsec:5and6}, and that of the addition of a site into a chain in Section~\ref{subsec:evenodd}, and explore the spectrum in terms of the invariant in Section~\ref{subsec:plotsandInvariant}, with concluding remarks in Section~\ref{sec:Conclusion}.

\section{Laser field coupled to the spin chain}
\label{sec:Hamiltonian}
The interaction part, $\widehat{H},$ of the Hamiltonian of our system, which is one mode (one frequency) of a laser field focused on one of the spins of the spin $\tfrac12$ chain of $N$ sites, reads:
\begin{equation} 
\widehat{H}=  G\left(a^{}S_k^{+} + a^\dag S^-_k \right)-
\sum_{i \neq j}2J_{i,j}\left(S_i^xS_{j}^x+S_i^y S_{j}^y\right).
\label{Eq:HamOCOperator}
\end{equation}
The $G$ part is the Jaynes-Cummings (JC) model of the photon ($a, a^\dag$) interacting with a two level system (spin $S_k$ on site $k$), and the second term (factor of 2 is there for convenience) is the XY model of the spin ($S_i$)-spin ($S_j$) interaction. We have in mind either nearest-neighbour (NN) case with $J_{ij}=J \delta_{r_j,r_i\pm d},$ or the long-range interactions in the form $J_{ij}= J |r_i-r_j|^{-2},$ with $J$ setting the scale for the spin interaction, $r_i, r_j$ being the coordinates of the spins on the lattice of nearest neighbour distance $d,$ and the indexes $i \textrm{ and } j$ enumerate the sites of the chain. For this report we have modeled an open chain geometry, where the ends do not influence each other, and a cyclic arrangement, a spin ring, both with either a NN or a long range spin-spin interaction. For a more detailed description of the notations and the meaning of the terms, as well as the choice of the basis vectors to form the matrix representation of this Hamiltonian, please refer to~\cite{Tonchev:2016JPCS}. 

\begin{figure}[th!!]\centering
\includegraphics[scale=0.35]{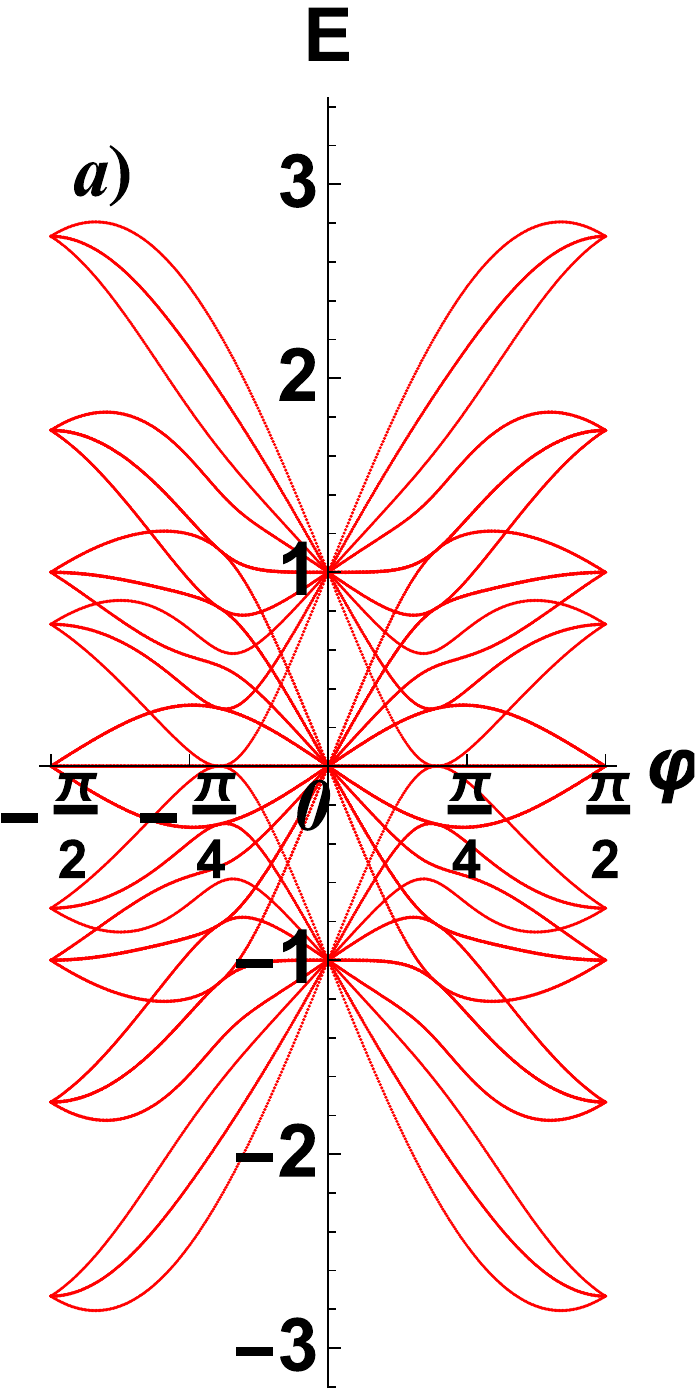}
\includegraphics[scale=0.35]{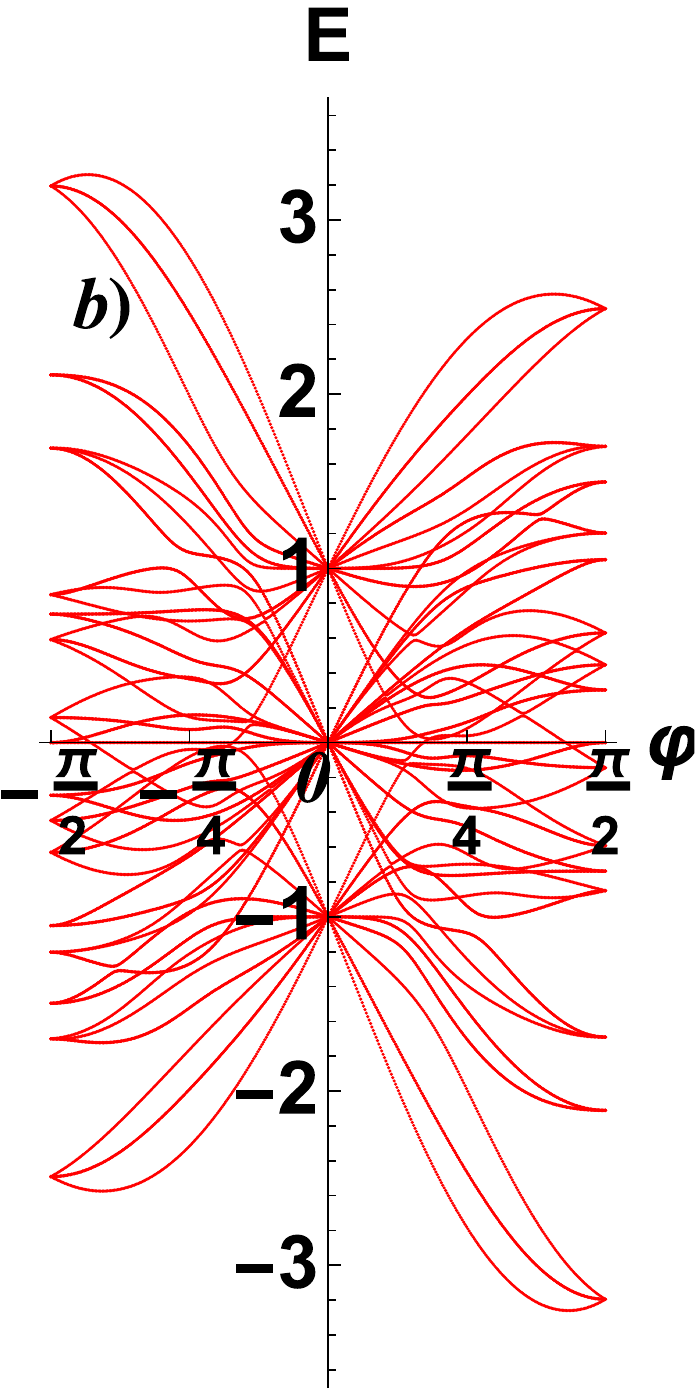} 
\hspace{0.1cm}\vrule\vrule\hspace{0.1cm}
\includegraphics[scale=0.35]{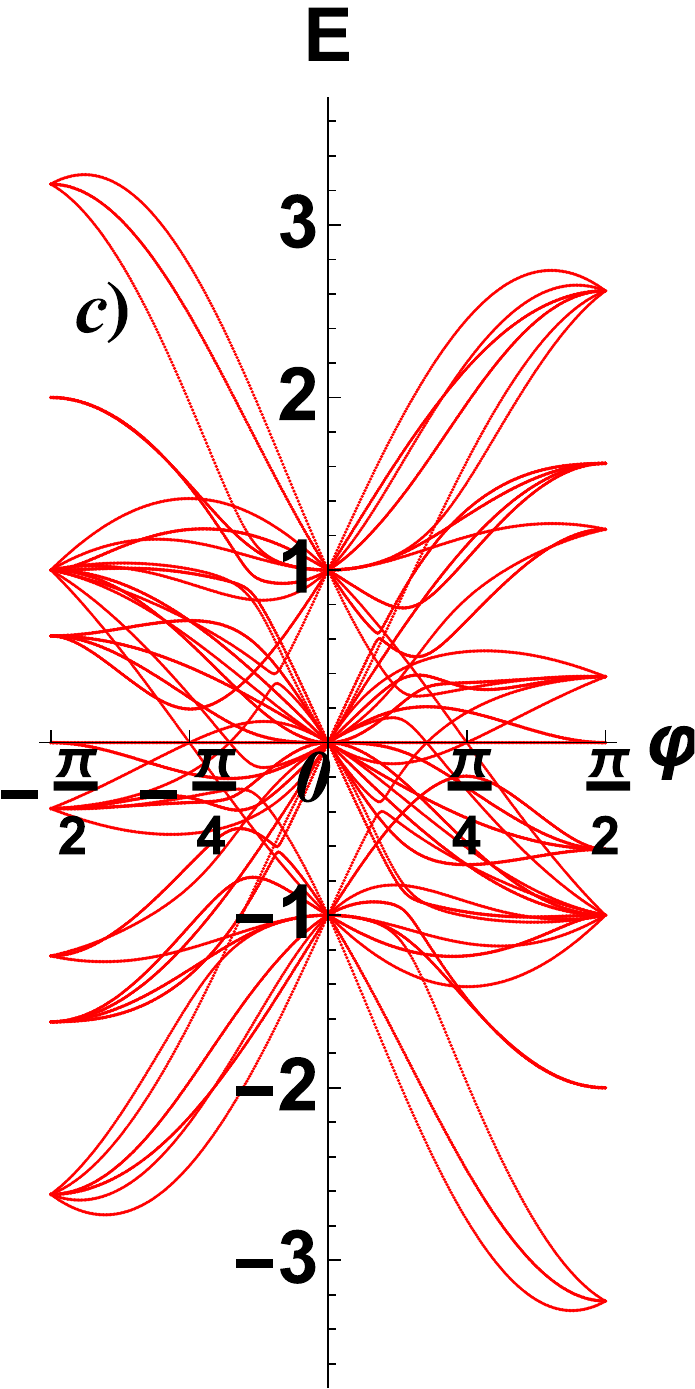} 
\includegraphics[scale=0.35]{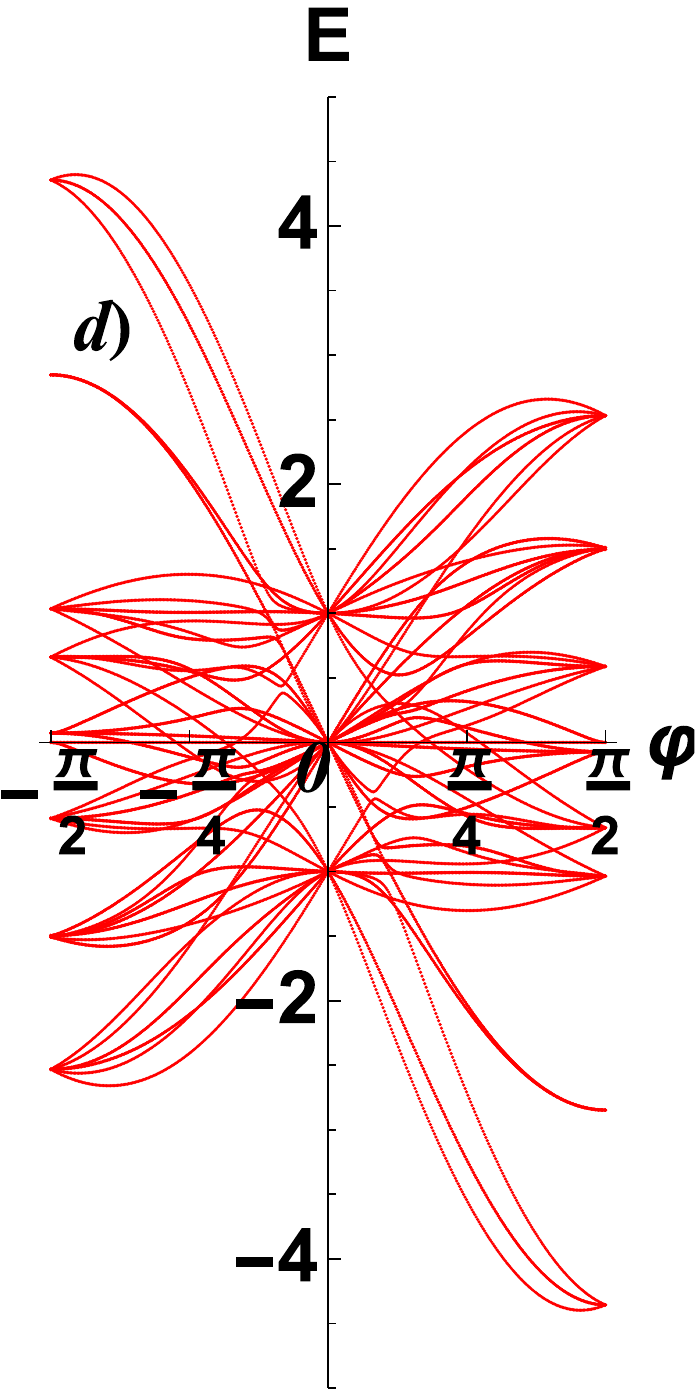}
\includegraphics[scale=0.35]{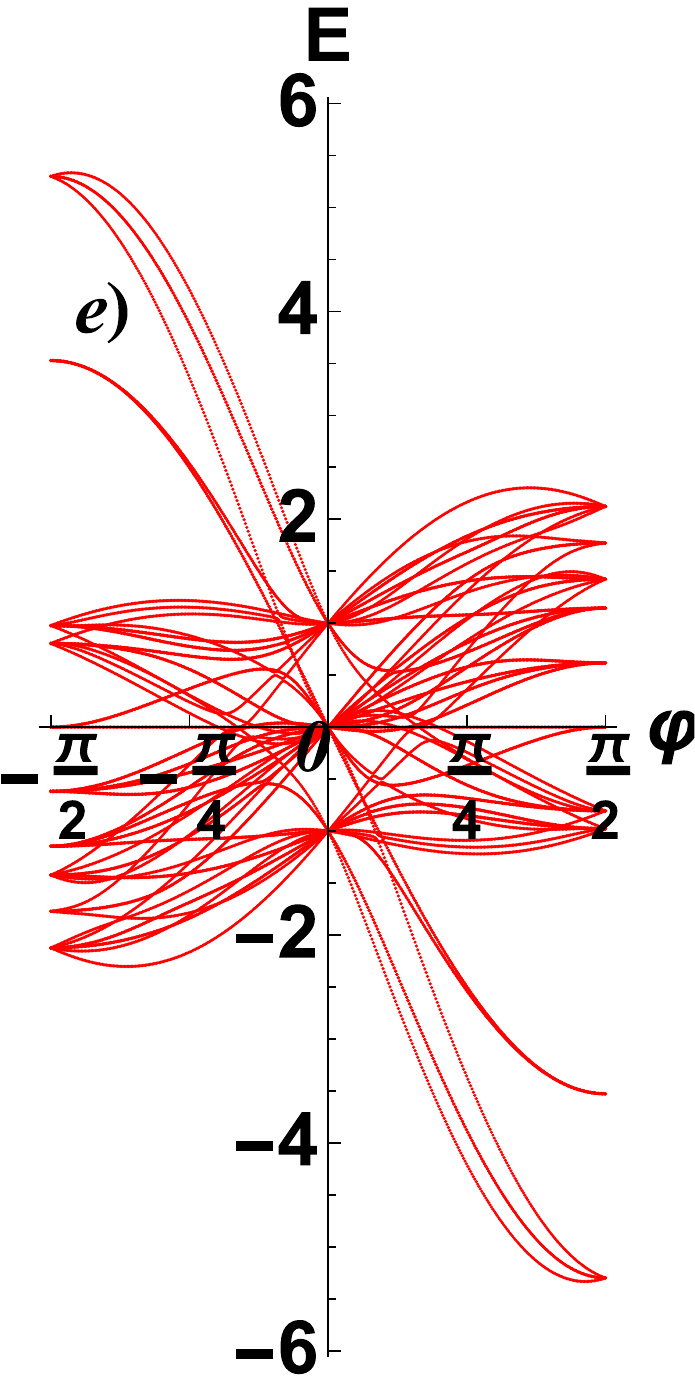} 
\\ 
\includegraphics[scale=.35]{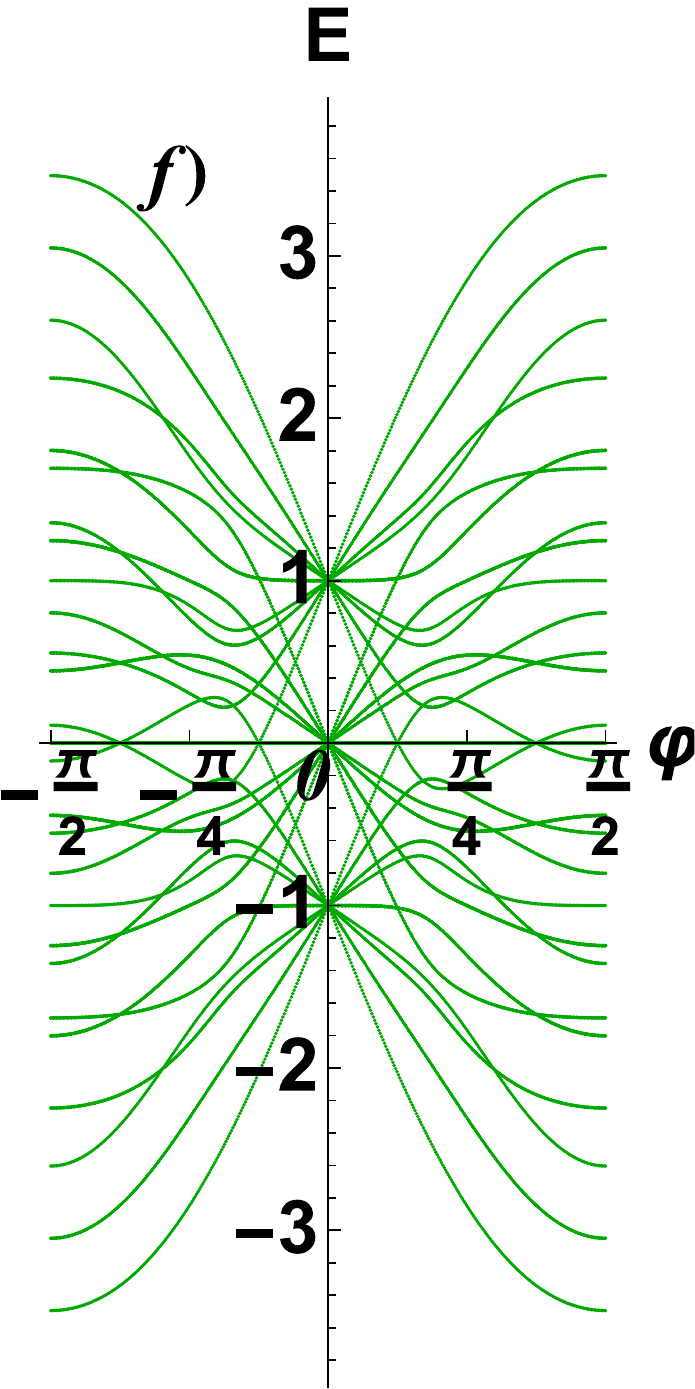}
\includegraphics[scale=0.35]{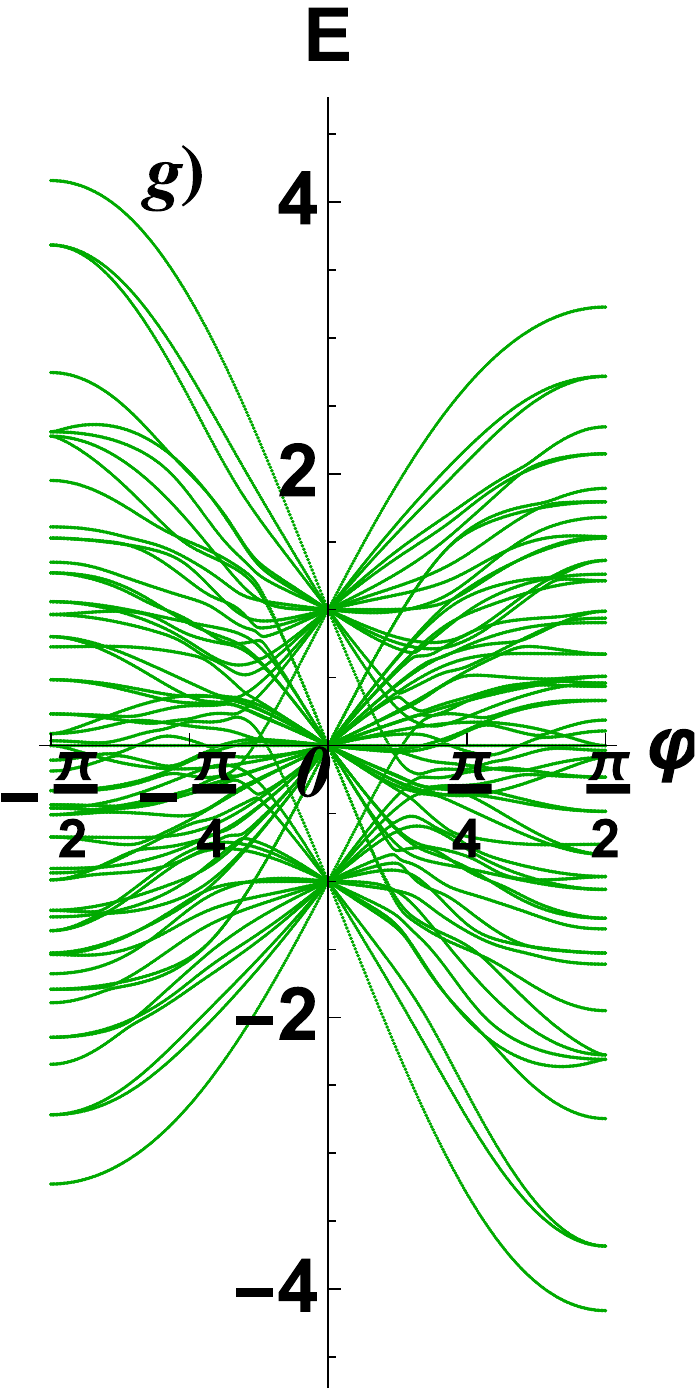}
\hspace{0.2cm}\vrule\vrule\hspace{0.2cm}
\includegraphics[scale=0.35]{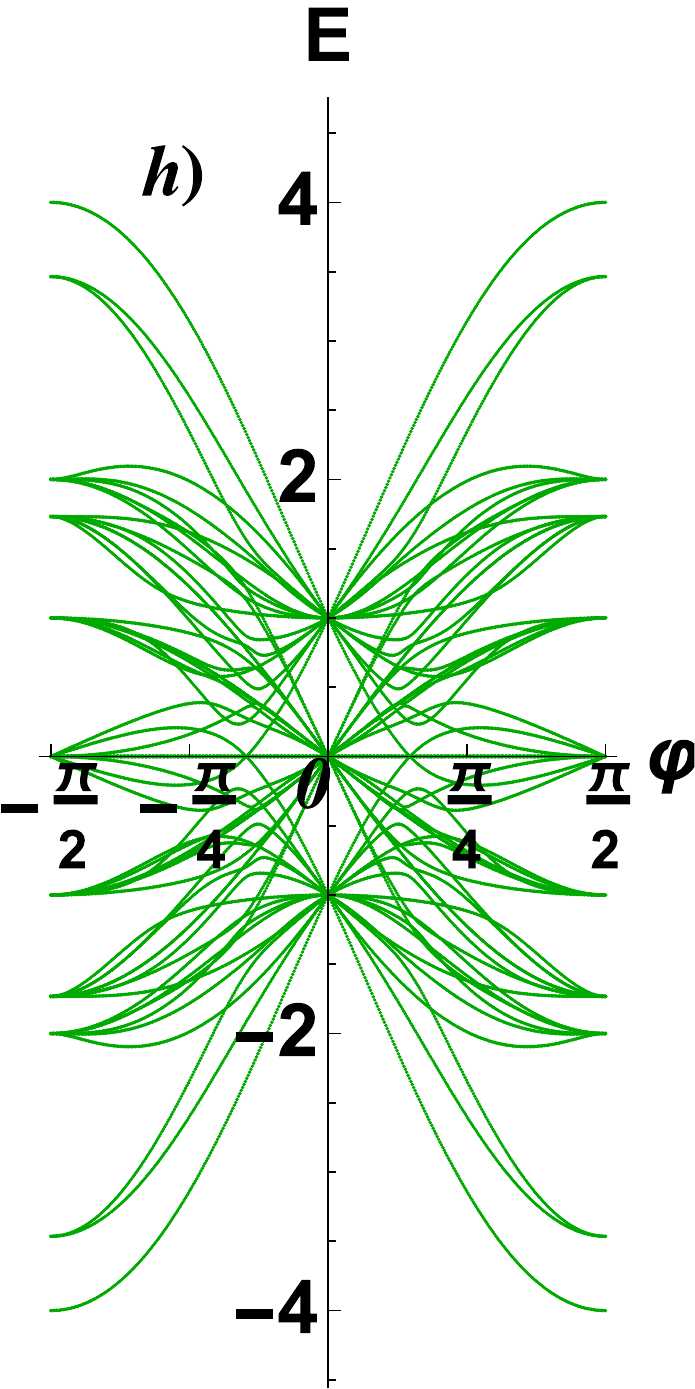}
\includegraphics[scale=0.35]{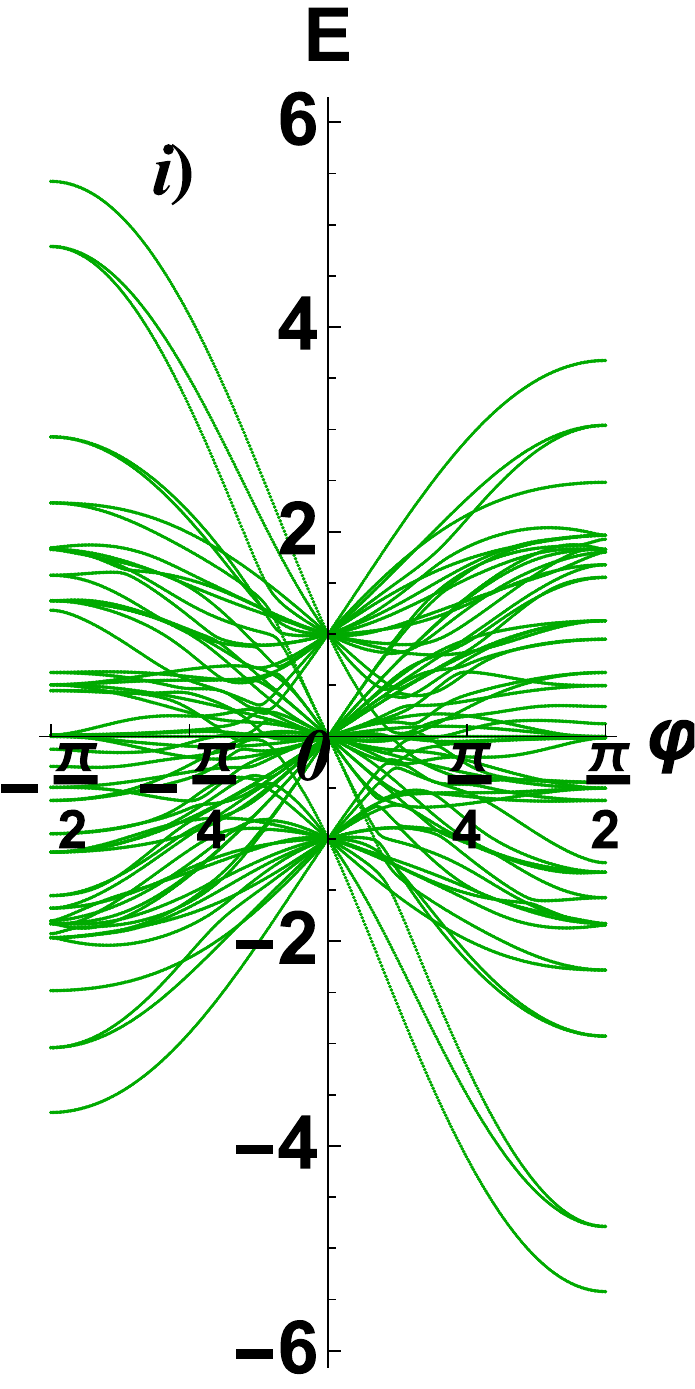}
\includegraphics[scale=0.35]{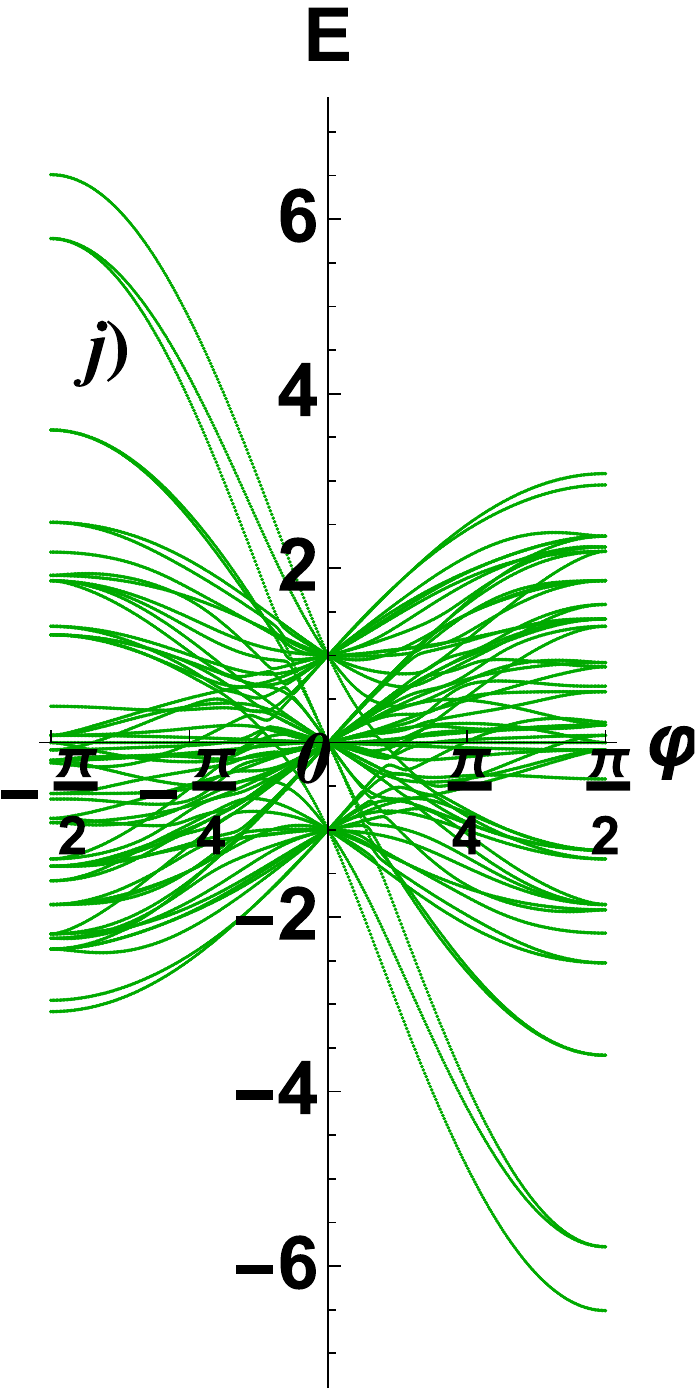}
\caption{(color online) The energy spectra, in units of $\sqrt{G^2+J^2},$ of the one photon JC-XY model for a 5-spin (top) and a 6-spin (bottom) open chain (two leftmost columns) with a NN (a,f) and long-range (b,g) spin-spin interaction, and for a spin ring arrangement (three rightmost columns), (c,h) NN spin-spin interaction; (d, i) long-range interaction taken along the ring and (e,j) long-range interaction considered to be along the hordes of the spin ring.
}
\label{Fig:ExamplesFor5and6}
\end{figure}

Solving the eigenvalue equation $\det (\widehat{H}-\lambda \textrm{ Identity})=0$ for a $2^{N+1}\times2^{N+1}$ matrix gives the energy spectra. These are shown in case of $N=5$ (top row) and $6$ (bottom row) for an open chain (first two columns) and for a spin ring (last three columns) in figure~\ref{Fig:ExamplesFor5and6}. We plot the values of $E(\varphi) = \lambda / \sqrt{G^2+J^2},$ for a cutout from the parameter space of $G$ and $J,$ such that $G^2+J^2 =1,\;\;G>0,$ with the parameter $\varphi$ defined as $J/G = \tan(\varphi).$ The pure magnetic case of $G=0$ corresponds to $\varphi=\pi/2,$ while the pure JC model falls on $\varphi=0$ as $J=0.$ The plots for $G<0$ are simply obtained from the figures by realizing that there is an overall change of sign in the energy, as seen from the equation~(\ref{Eq:HamOCOperator}), when $G\rightarrow-G,J\rightarrow-J$ then $ E\rightarrow-E.$ The total number of energy values for each $\varphi$ scales as $2^{N+1},$ consequently, even with the degeneracy taken into account, the figures for $N=9$ and $10$ become too messy, but show a similar behaviour. Finally, the resulting energy spectra represent the mixture of the photon and the spins, while for an easier comparison with the known results, for example for the a two-level atom with a Jaynes-Cummings model, we will need to project away the photon.

\section{Analysis of the energy spectra}
\label{sec:analyze}
\vspace{\baselineskip}
\subsection{Magnetization and the Invariant for a one photon problem}
\label{subsec:Invariant}
As it will be seen from the analysis of the plots it is helpful to have an understanding of the magnetic part of the problem, at $\varphi=\tfrac{\pi}2.$  First of all, from the two physical entities, the magnetic moment $M_z=\frac{1}{2}\sum_{i=1}^N\sigma_i^{z},$ and the photon fill number, $n_{ph}=a^{\dag}a,$ we can create an invariant, a combination that commutes with the Hamiltonian of the JC+XY model~\cite{RJZSHM2015}: $Inv=n_{ph}+M_z,\; [Inv,\widehat{H}]=0.$

In~\cite{Tonchev:2016JPCS} the states for the electromagnetic field were ordered by first choosing the fill number zero $| 0 \rangle,$ then the fill number one $|1\rangle$ state. If instead we choose first $|1\rangle, $ and then $|0\rangle,$ while keeping the same order for the spin basis on each site: $|\uparrow\rangle,|\downarrow\rangle$, then this "one" photon problem is represented in terms of spin variables. In this basis the matrix form, in terms of the Pauli matrices, of the photon fill number is similar to introducing the isotopic spin in describing the two states of a nucleon, a neutron and the proton, see e.g. \S116 in~\cite{LandauLifshitzIII}:
\begin{equation}
n_{ph} \equiv \frac12+\frac12 \sigma_z,
\end{equation}
the JC interaction [$G$ part in equation~(\ref{Eq:HamOCOperator})] is effectively a XY type interaction, between site K and the extra site $0$ chosen to correspond to the photon:
\begin{equation}
a^{}S_k^{+} + a^\dag S^-_k  \equiv \sigma^{-}_{\textrm{site }0}S_k^{+} + \sigma^{+}_{\textrm{site }0}S_k^{-},
\end{equation}
and, finally, the invariant is the sum over the chain sites of a spin-1/2 chain with one extra site $i=0:$
\begin{equation}
Inv = \underbrace{\frac{1}{2}+\frac{\sigma_z}{2}}_{\textrm{site }0} + M_z = \frac{1}{2}+ \mathcal{M}_z, \;\; \textrm{where } \mathcal{M}_z\overset{def}{=}\frac{1}{2}\sum_{i=0}^N\sigma_{i}^{z}.
\end{equation}
We thus have the eigenvalues of $Inv:$ they are immediately obtained by analogy with the familiar problem for the $z$-component of the angular momentum as $-(N+1)/2,-(N+1)/2+1,\ldots, (N+1)/2,$ only shifted by $\tfrac12.$

We have a possible mapping of the problem of $N$ spins and a photon onto a pure spin chain of $N+1$ sites and an anisotropic interaction. 
Numerically, we can benefit from the advanced packages for solving numerical spin chain problems.

Another useful observation is that for the XY interaction the states with maximum absolute energy are those that have the most number of opposite spin pairs. For example, the $\left|\uparrow \downarrow \rangle \right.$ and $\left|\downarrow \uparrow \rangle \right.$  pairs with $M_z=0$ contribute $\pm J,$ and for $J>0$ the antisymmetric combination of such AFM states combines to give the highest energy level. For $3-$spins similarly composed product states $\left|\uparrow \downarrow \uparrow\rangle \right.$  or $\left| \downarrow \uparrow \downarrow\rangle \right.$  states have $M_z=\pm \tfrac12.$ In case $J<0$ the symmetric combinations combine for the highest energy level. The fully polarized spin pairs, the FM states $|\uparrow \uparrow \rangle$ with maximum values of $M_z, $ contribute zero, $0 J,$ to the energy. 

For the one photon problem, the presence of the photon states, even at $G=0,$ gives by analogy, that the states with maximal $\mathcal{M}_z,$ $|1_{ph} \uparrow \uparrow \rangle \ldots$ and $|0_{ph}\downarrow \downarrow \rangle \ldots,$ are the states with an identically equal to zero energy for any $G$ and $J$ values, while the states with $\mathcal{M}_z=0$ (odd number $N$) or $\mathcal{M}_z=\pm \tfrac12$ (even number $N$) have the maximum energy, which is to be compared with $M_z=0$ (even number $N$) or $M_z=\pm \tfrac12$ (odd number $N$) from above.

\subsection{NN vs long range interaction}
\label{subsec:5and6}
\textit{Open chain:} 
The comparison between the first two columns in figure~\ref{Fig:ExamplesFor5and6} illustrates the influence of the range of the spin-spin interaction, NN in the first column or long range in the second. The spectrum in figure~\ref{Fig:ExamplesFor5and6}a) is symmetric for reflections about the x and the y axes, and this means, in terms of the interaction, that the NN interaction gives energies that are independent of the sign of either of the parameters $G \textrm{ or } J,$ $E^{(a)}(G,J)=E^{(a)}(-G,J)=E^{(a)}(G,-J).$ 
The spectrum in figure~\ref{Fig:ExamplesFor5and6}b) continues to be independent of the overall sign of the JC part $G,$ as the first and third quadrant energy bands, as well as the 2nd and the 4th, are symmetric to each other, but dependents on the sign of the spin-spin interaction $J,$ as the first and the second quadrant plots are different, $E^{(b)}(G,J)=E^{(b)}(-G,J) \neq E^{(b)}(G,-J)$. 

\textit{Spin ring:} Looking at the last three columns figure~\ref{Fig:ExamplesFor5and6}c), d), and e) we can no longer claim that the spectrum in NN case is independent of the signs of $G$ and $J,$ as this depends on whether the spin ring has an odd (top row),where this is no longer the case, or an even (bottom row) number of spins, for which this is similar to the open chain. 
This follows from the fact that the XY eigenvalues for the spin ring no longer cluster symmetrically around the zero, $E^{ring} (\pi/2) \neq - E^{ring} (\pi/2),$ even though their weighted average, with the degeneracies taken into account, still is zero. 
This translates into the 2nd quadrant, by making the spectrum loose its symmetry $E(G,J)\neq E(G,-J).$

\subsection{Even vs Odd number of spin sites}
\label{subsec:evenodd}
Now we compare the top and bottom row of the figures~\ref{Fig:ExamplesFor5and6} in order to analyze the effect of introducing an extra spin site into the system. 
The most striking feature is the more effective lifting of the degeneracy in the odd site spin chain when we move from the purely magnetic case at $\varphi=\pi/2.$ Another feature is that the maximum energy values are reached for even $N$ at this magnetic boundary, while for the odd $N$ the maximum is always achieved when a photon is mixed as well, reaching the magnetic boundary only at $N_{odd} \rightarrow \infty.$

The more effective lifting of the degeneracy in the case of odd $N$ can be understood when we consider the effect of the JC interaction applied to eigenstates of the problem at $\phi=\pi/2.$ In a poor man's perturbation approach, we look only at the maximum energy state. As mentioned at the end of  Section~\ref{subsec:Invariant} the corresponding eigenstates are a mixture of the type $|\downarrow \uparrow \downarrow \rangle \ldots.$ Then, for example, if the photon interacts with the first spin, the JC part of the Hamiltonian has amplitudes of transition between this and a state written as $|\uparrow \uparrow \downarrow \rangle \ldots,$ which means that it has to be diagonalized using more states, than at the point $\phi=\pi/2,$ and this bigger matrix has more eigenvalues, thus lifting the degeneracy. For the even numbered, for example $N=4,$ we have the maximum energy state at $\phi=\pi/2$ as a combination of $|\downarrow \uparrow \downarrow \uparrow \rangle,$ and the JC part would have matrix elements with smaller range of states and thus there is no lifting of the degeneracy.

\subsection{Some spectroscopic analysis using the invariant.}
\label{subsec:plotsandInvariant}
\begin{figure}[ht!]\centering
\includegraphics[scale=.4]{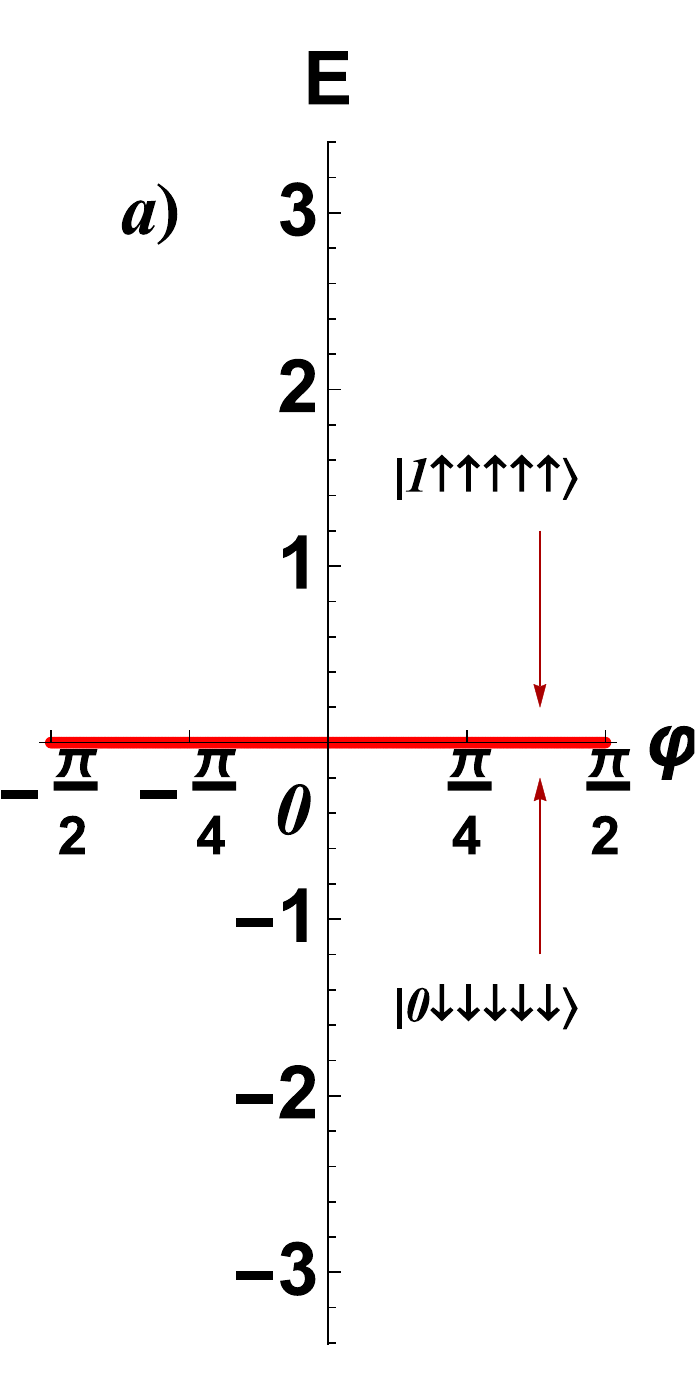}
\includegraphics[scale=.4]{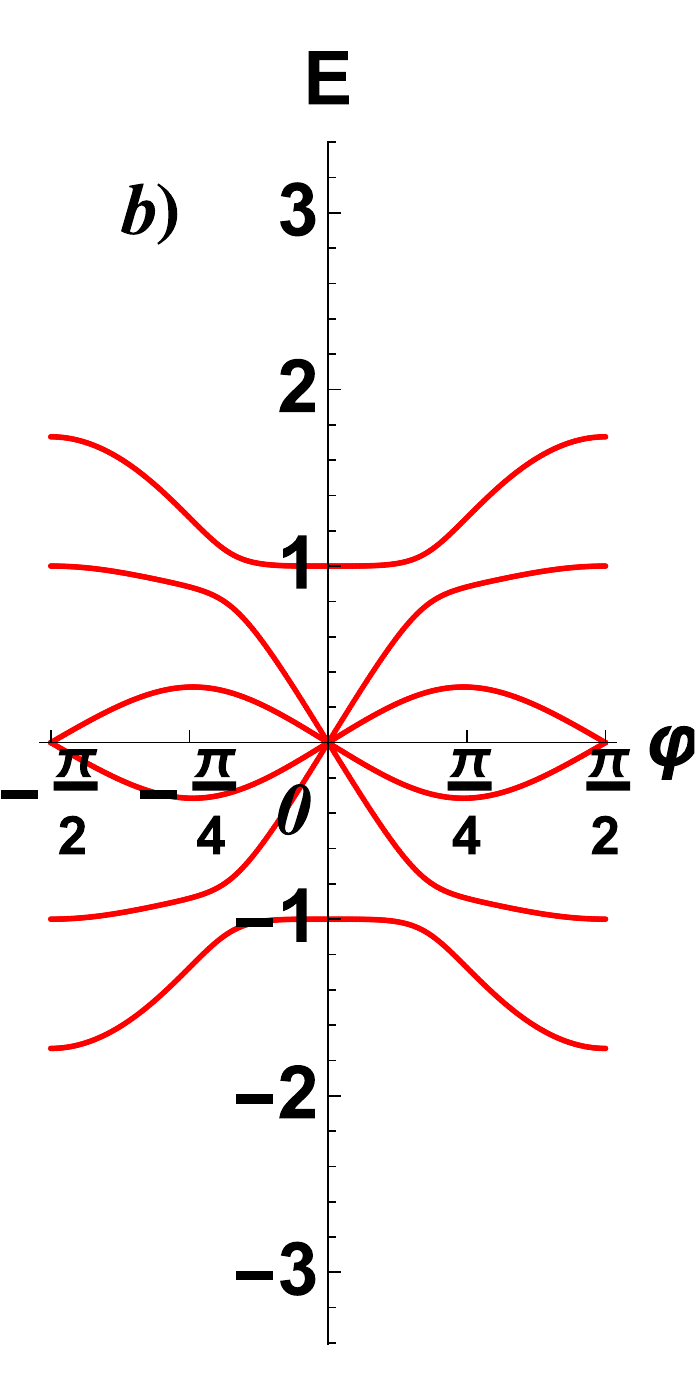}
\includegraphics[scale=.4]{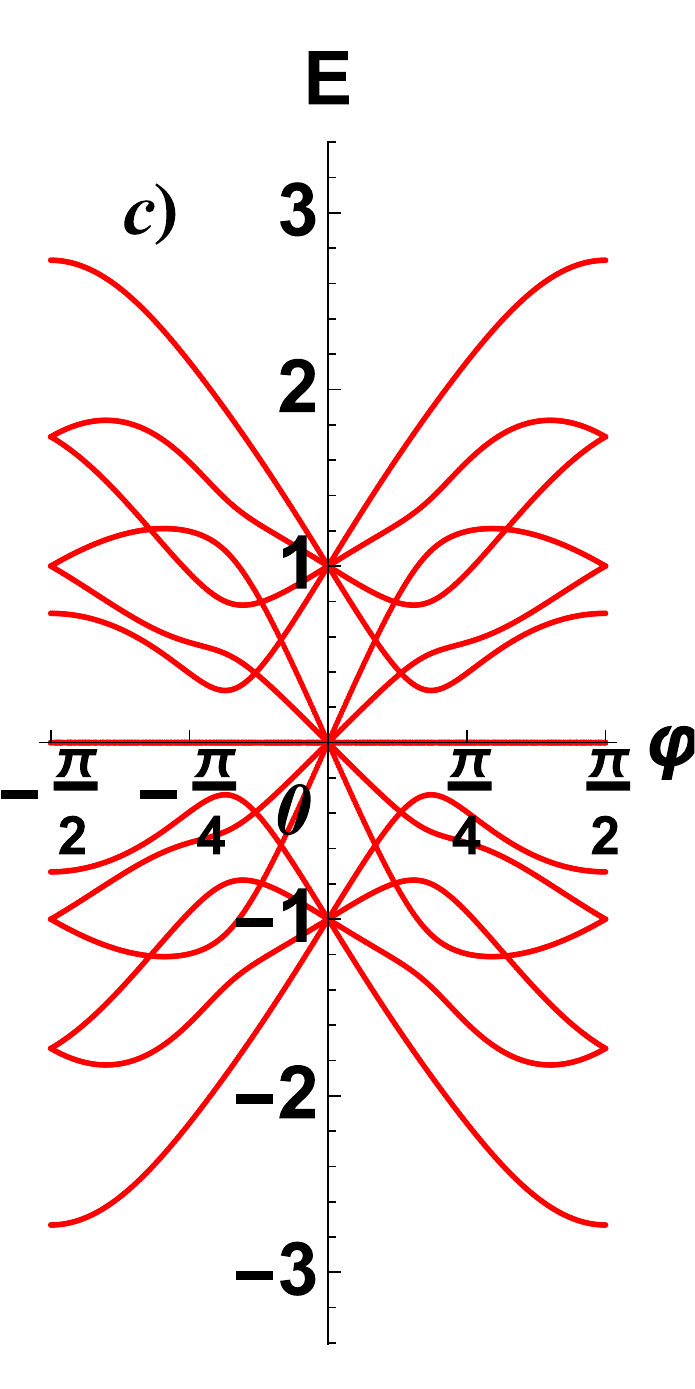} 
\includegraphics[scale=.4]{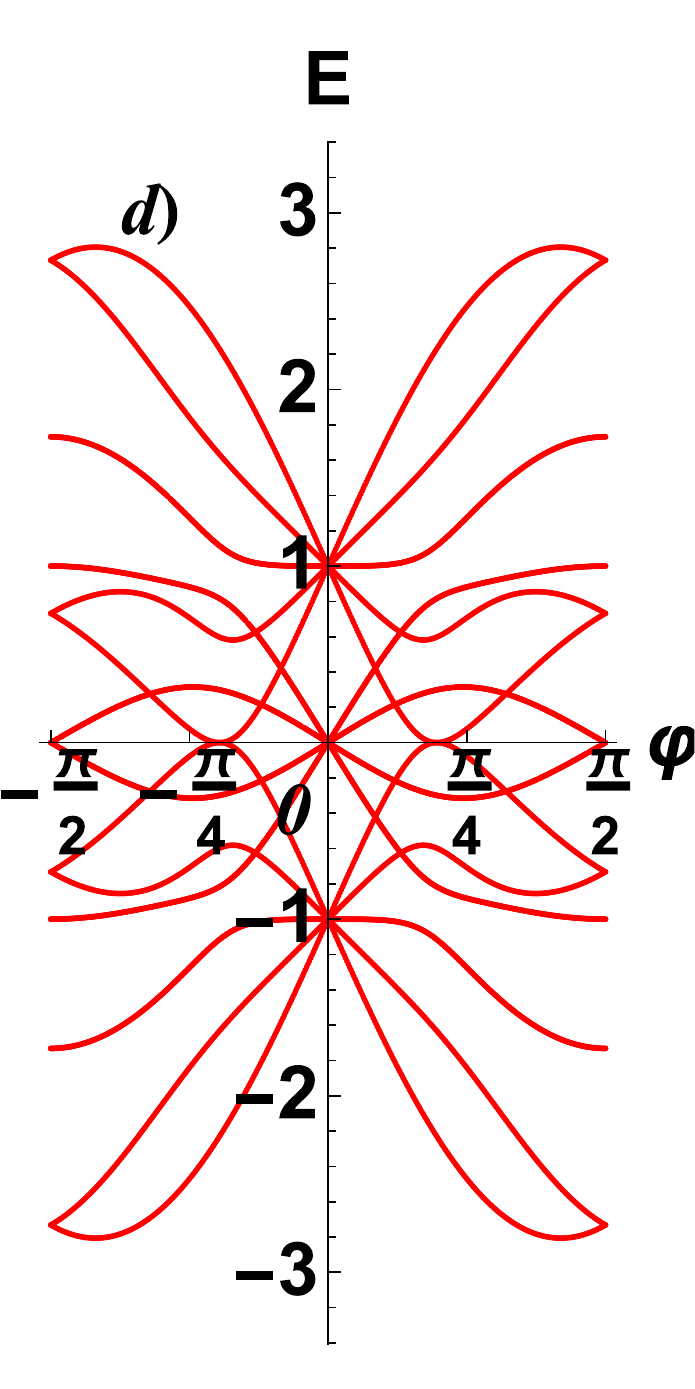}\\
\includegraphics[scale=.4]{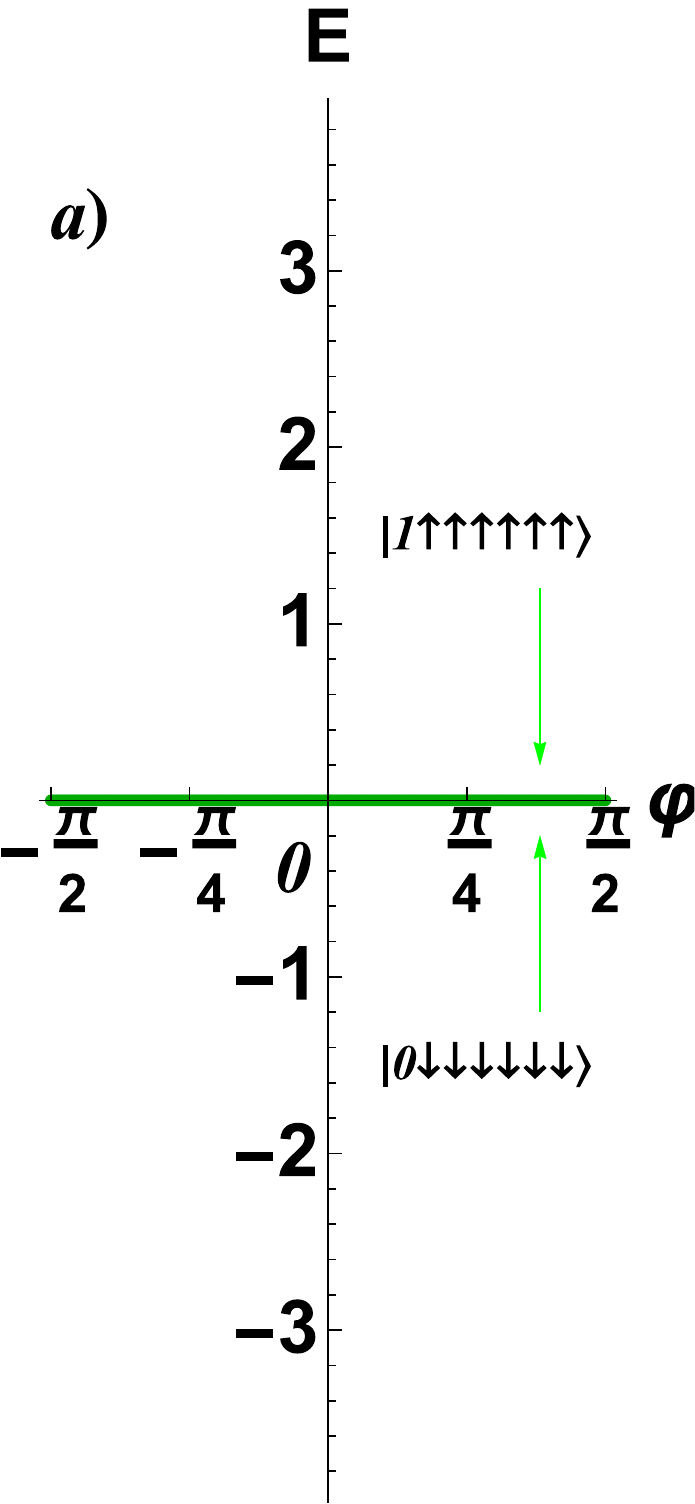}
\includegraphics[scale=.4]{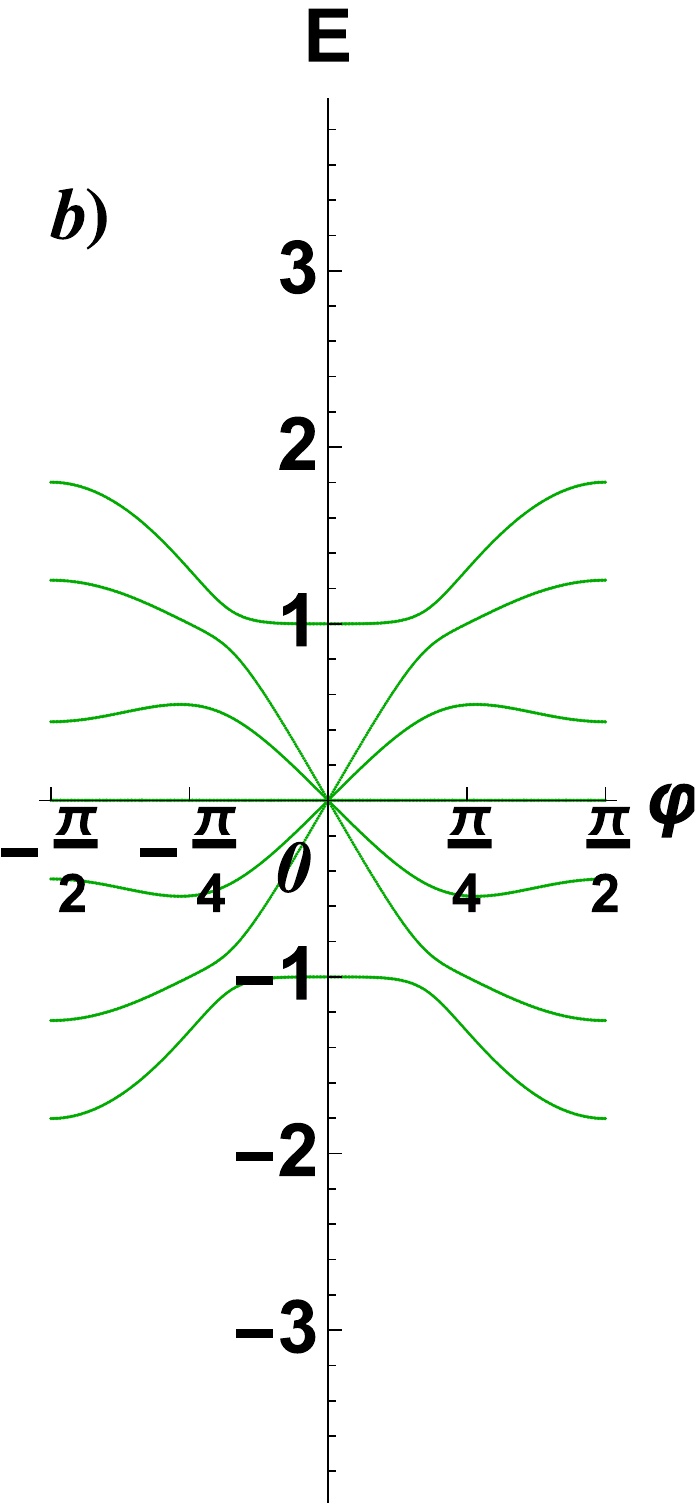}
\includegraphics[scale=.4]{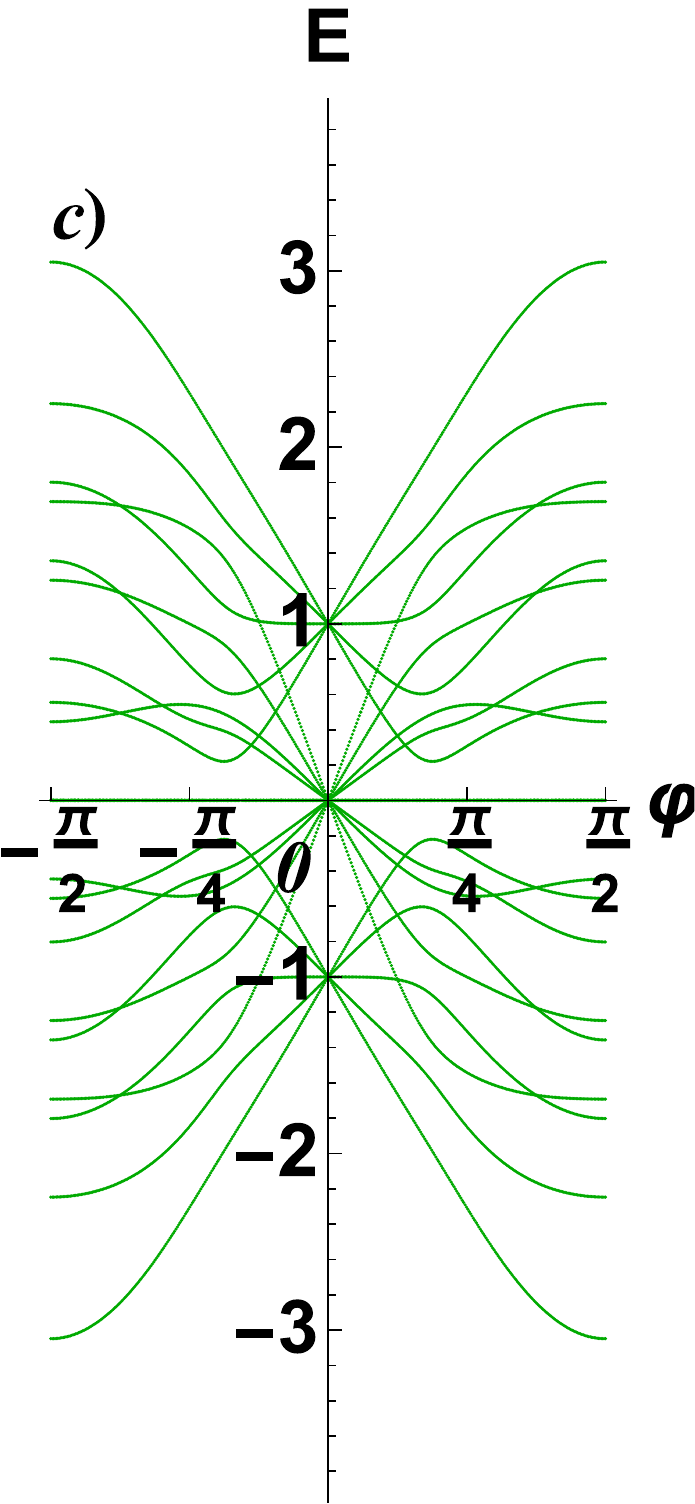} 
\includegraphics[scale=.4]{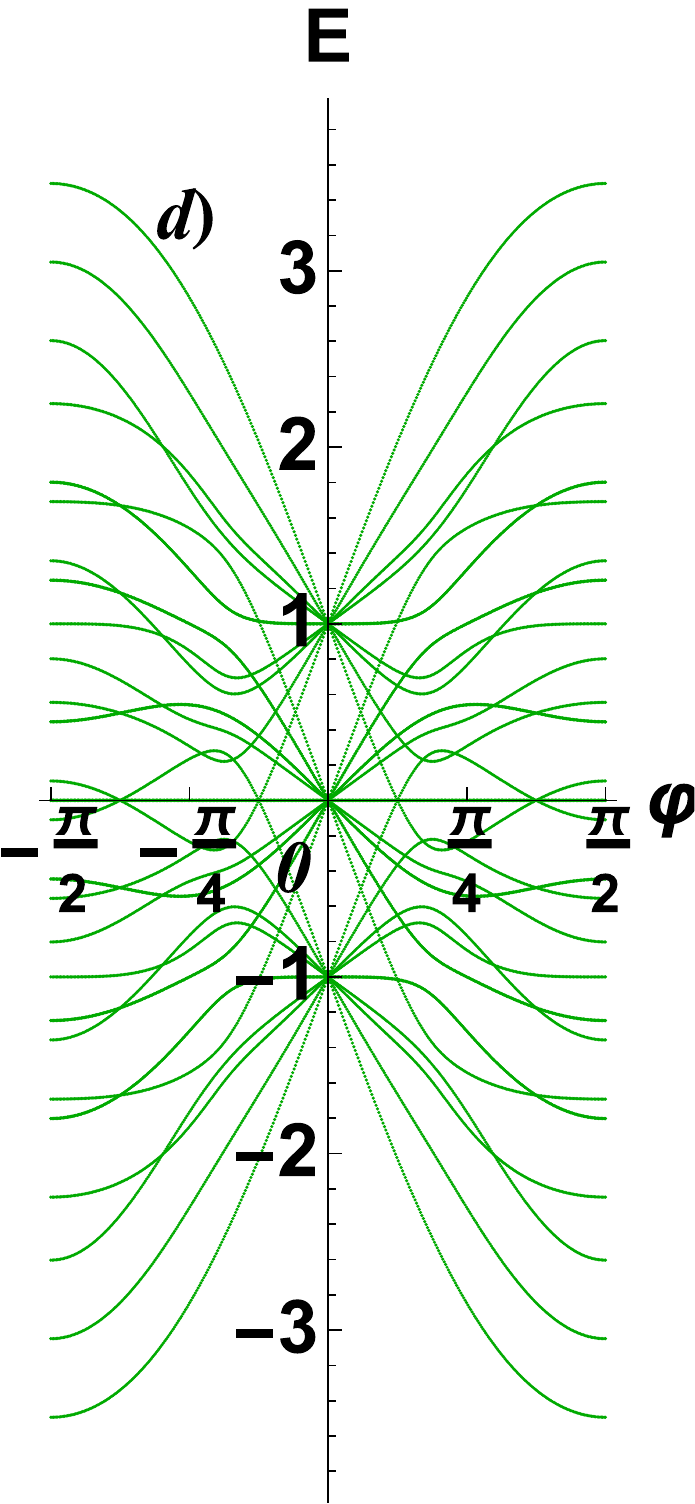}
\caption{(color online) Energy levels sorted by the invariant for the test case of an open spin chain with NN interaction. 5 spins (top row) or 6 spins (bottom row).
Levels with $|\mathcal{M}_z| $ equal to: (a) $3\; (N=5)$ and $7/2\; (N=6)$  (b) $2\; (N=5)$ and $5/2 \;(N=6)$ (c) $1\; (N=5)$ and $3/2 \;(N=6)$ $0\; (N=5)$ and $1/2\; (N=6)$
}
\label{Fig:invariantspectroscopy5and6}
\end{figure}

The values of the invariant, in terms of $Inv$ or $\mathcal{M}_z,$ can be used to select bands with particular property. For example, let's consider the test case of an open spin chain with 5 or 6 spins with the NN interaction. For $N=5,$ $\mathcal{M}_z = -3,-2,...,2,3.$  Similarly, for $N=6$ we get $\mathcal{M}_z  = -7/2,\ldots, 7/2.$ Some of the graphs for this case are given in figure~\ref{Fig:invariantspectroscopy5and6}. For $|\mathcal{M}^{(N=5)}_z|=3,$ and $|\mathcal{M}^{(N=6)}_z|=7/2,$ the energy levels are identically equal to zero, and we show the typical eigenstate. 
In Table~1 we give some of  the numeric details for this open chain case.

\begin{table}[ht!]\centering
\begin{tabular}{ |c|c|c|c| }\hline
$N$ & $\#$ of Distinct & $E_{max}$ & $\varphi_{max}$  \\ \hline 
2 & $3$      & 1       & flat \\  
3 & $3$     & 1,6180 & -1,0196  \\
4 & $9$     & 2,2361 & -1,5708  \\
5 & $9$    & 2,8064 & -1,3188  \\
6 & $27$    & 3,4940 & -1,5708  \\
7 & $27$    & 4,0649 & -1,4212  \\
8 & $58$    & 4,7588 & -1,5708 \\
9 & $91$        & 5,3362 & -1,4684  \\
10& $256$     & 6,0267  & -1,5708 \\ \hline
\end{tabular}
\label{Tab:invariantspectroscopy6}
\caption {Summary of the number of distinct eigenvalues at $\varphi=\pi/2$, the maximum value of the energy and the location of this maximum for each of the nine lengths $N$ of an open spin chain molecule with NN interaction.}
\end{table}

\newpage
\section{Conclusion}
\label{sec:Conclusion}
In this report we have presented numerical results for one photon Jaynes-Cummings XY model for a spin ring or an open spin molecules of sies of up to 10 spin-$\tfrac12$'s. The resulting energy spectra represent the mixture of the photon and the spins, while for an easier comparison with the known results, for example for the a two-level atom with a Jaynes-Cummings model, we will need to project away the photon. Some improvement of the total number of included spins may achievable by using advanced spin chain simulation packages, because the one photon (i.e., the photon filling number) approximation restricts the problem into a spin type chain. Another useful case, with filling number being in the range $>1$ is a subject of future work. 

\ack
AAD would like to thank prof.~J.~Schnack (UniBielefeld) for hospitality, where part of this work was completed, and prof.~N.~Tonchev (ISSP, BAS) for useful comments. 
This work was supported by EU FP7 INERA project grant agreement number 316309.
 
\section*{References}
\bibliographystyle{iopart-num}
\bibliography{INERA-VPT-2016-07-IOP}\
\end{document}